\documentstyle[12pt,aaspp4]{article}
\newread\epsffilein    
\newif\ifepsffileok    
\newif\ifepsfbbfound   
\newif\ifepsfverbose   
\newdimen\epsfxsize    
\newdimen\epsfysize    
\newdimen\epsftsize    
\newdimen\epsfrsize    
\newdimen\epsftmp      
\newdimen\pspoints     
\def\epsfbox#1{\global\def\epsfllx{72}\global\def\epsflly{72}%
   \global\def\epsfurx{540}\global\def\epsfury{720}%
   \def\lbracket{[}\def\testit{#1}\ifx\testit\lbracket
   \let\next=\epsfgetlitbb\else\let\next=\epsfnormal\fi\next{#1}}%
\def\epsfgetlitbb#1#2 #3 #4 #5]#6{\epsfgrab #2 #3 #4 #5 .\\%
   \epsfsetgraph{#6}}%
\def\epsfnormal#1{\epsfgetbb{#1}\epsfsetgraph{#1}}%
\def\epsfgetbb#1{%
\openin\epsffilein=#1
\ifeof\epsffilein\errmessage{I couldn't open #1, will ignore it}\else
   {\epsffileoktrue \chardef\other=12
    \def\do##1{\catcode`##1=\other}\dospecials \catcode`\ =10
    \loop
       \read\epsffilein to \epsffileline
       \ifeof\epsffilein\epsffileokfalse\else
          \expandafter\epsfaux\epsffileline:. \\%
       \fi
   \ifepsffileok\repeat
   \ifepsfbbfound\else
    \ifepsfverbose\message{No bounding box comment in #1; using defaults}\fi\fi
   }\closein\epsffilein\fi}%
\def\epsfsetgraph#1{%
   \epsfrsize=\epsfury\pspoints
   \advance\epsfrsize by-\epsflly\pspoints
   \epsftsize=\epsfurx\pspoints
   \advance\epsftsize by-\epsfllx\pspoints
   \epsfxsize\epsfsize\epsftsize\epsfrsize
   \ifnum\epsfxsize=0 \ifnum\epsfysize=0
      \epsfxsize=\epsftsize \epsfysize=\epsfrsize
     \else\epsftmp=\epsftsize \divide\epsftmp\epsfrsize
       \epsfxsize=\epsfysize \multiply\epsfxsize\epsftmp
       \multiply\epsftmp\epsfrsize \advance\epsftsize-\epsftmp
       \epsftmp=\epsfysize
       \loop \advance\epsftsize\epsftsize \divide\epsftmp 2
       \ifnum\epsftmp>0
          \ifnum\epsftsize<\epsfrsize\else
             \advance\epsftsize-\epsfrsize \advance\epsfxsize\epsftmp \fi
       \repeat
     \fi
   \else\epsftmp=\epsfrsize \divide\epsftmp\epsftsize
     \epsfysize=\epsfxsize \multiply\epsfysize\epsftmp   
     \multiply\epsftmp\epsftsize \advance\epsfrsize-\epsftmp
     \epsftmp=\epsfxsize
     \loop \advance\epsfrsize\epsfrsize \divide\epsftmp 2
     \ifnum\epsftmp>0
        \ifnum\epsfrsize<\epsftsize\else
           \advance\epsfrsize-\epsftsize \advance\epsfysize\epsftmp \fi
     \repeat     
   \fi
   \ifepsfverbose\message{#1: width=\the\epsfxsize, height=\the\epsfysize}\fi
   \epsftmp=10\epsfxsize \divide\epsftmp\pspoints
   \vbox to\epsfysize{\vfil\hbox to\epsfxsize{%
      \includegraphics{#1}%
      \hfil}}%
\epsfxsize=0pt\epsfysize=0pt}%

{\catcode`\%=12 \global\let\epsfpercent=
\long\def\epsfaux#1#2:#3\\{\ifx#1\epsfpercent
   \def\testit{#2}\ifx\testit\epsfbblit
      \epsfgrab #3 . . . \\%
      \epsffileokfalse
      \global\epsfbbfoundtrue
   \fi\else\ifx#1\par\else\epsffileokfalse\fi\fi}%
\def\epsfgrab #1 #2 #3 #4 #5\\{%
   \global\def\epsfllx{#1}\ifx\epsfllx\empty
      \epsfgrab #2 #3 #4 #5 .\\\else
   \global\def\epsflly{#2}%
   \global\def\epsfurx{#3}\global\def\epsfury{#4}\fi}%
\def\epsfsize#1#2{\epsfxsize}
\let\epsffile=\epsfbox

\def\ni{\noindent}

\def\beq{\begin{equation}}
\def\ee{\end{equation}}
\def\lsim{\mathrel{\rlap{\lower4pt\hbox{\hskip1pt$\sim$}}
    \raise1pt\hbox{$<$}}}
\def\gsim{\mathrel{\rlap{\lower4pt\hbox{\hskip1pt$\sim$}}
    \raise1pt\hbox{$>$}}}

\def\bfJ{{\bf J}}

\def\bfA{{\bf A}}

\def\bfB{{\bf B}}

\def\ts{\times}
\def\lb{\langle}
\def\rb{\rangle}
\def\curl{\nabla {\ts}}

\def\bfv{{\bf v}}

\def\bfB{{\bf B}}

\begin{document}

\setcounter{equation}{0}

\title{How Spectral Shapes of Magnetic Energy and Magnetic Helicity 
Influence their Respective Decay Time Scales}



\medskip

\author{Eric G. Blackman} 
\affil{ Department of Physics \& Astronomy and Laboratory for
Laser Energetics, University of Rochester, Rochester NY 14627;
email:  blackman@pas.rochester.edu}

\centerline {(submitted to, Plasma Phys. and Cont. Fusion)} 

\begin{abstract}

It is shown that in  magnetohydrodynamics, 
the shapes of the magnetic energy and magnetic helicity 
spectra can influence whether the magnetic helicity resistive 
dissipation time scale is  
longer, equal to, or shorter than that of magnetic energy.
The calculations highlight that 
magnetic helicity need not always dissipate significantly more slowly 
than magnetic energy. 
While this may be implicitly understood in detailed studies 
of magnetic helicity in plasma devices, it is not sufficiently 
emphasized in elementary discussions comparing 
magnetic helicity vs. magnetic energy evolution.
Measuring magnetic energy and helicity decay times may provide 
{\it a posteriori} insight into the initial spectra.
\end{abstract}

\centerline{PACS codes: 52.30.Cv, 95.30.Qd, 
52.65Kj, 52.55 -s, 96.60.Hv, 98.62Mw}




\bigskip


\section{Introduction}

Magnetic helicity, (defined by $\int \bfA\cdot\bfB d^3x$,
where $\bf A$ is the vector potential, $\bf B$ is the magnetic field
and $\int d^3x$ is a volume integral) is 
a measure of the linkage and twist of magnetic
field lines, and is a conserved quantity for a closed system
in ideal magnetohydrodynamics [\cite{woltjer,berger}]. 
Magnetic helicity is often more 
strongly conserved than magnetic energy 
in the presence of resistive dissipation 
[\cite{berger}-\cite{bellan00}].
The concept of a slower decay of magnetic helicity than magnetic energy 
forms the basis for Taylor relaxation theory
[\cite{berger}-\cite{bellan00}], 
which describes the end state to which magnetically dominated
magnetohydrodynamic configurations relax.
The theory is well known in the context of laboratory pinches 
and has also been applied to the solar corona 
[\cite{heyvaertspriest84}] and astrophysical jets [\cite{koniglchoudhuri85}].
Taylor theory predicts the relaxed state  by minimizing the magnetic 
energy subject to the constraint that magnetic helicity is conserved.
The result is a force-free configuration
with the scale of the field reaching the largest gradient scale available 
subject to boundary conditions.  {Some systems (e.g. Reverse Field Pinches (RFP)) 
transit toward and away from the Taylor state 
[e.g. \cite{ortolani93}], when driven.

The relation between magnetic helicity and magnetic energy decay rates
has been studied in [\cite{berger2}] but without 
focusing on the role that energy and magnetic
helicity spectra play in their relative magnitudes.
Here I specifically 
show how the relative shapes of the magnetic and magnetic helicity
spectra can influence the ratio of magnetic energy to magnetic helicity
decay time scales. 
{I derive the spectral shapes of magnetic helicity
and magnetic energy for which the magnetic energy decays faster,
equal, or more slowly than magnetic helicity. 
The purpose of these simple calculations 
is to illustrate very simply 
how the commonly intuited slow decay of magnetic helicity
compared to magnetic energy 
is influenced by the spectral shapes of these quantities.

\section{Magnetic Helicity vs. Magnetic Energy Decay Time Scales}
In what follows, I will assume that surface terms are small
compared to other terms involved in the time evolution of the 
the magnetic energy, magnetic helicity, 
and kinetic energies of the system. This would be the case
if injection or loss of these quantities through the boundaries
proceeds on time scales which are long compared to that of the  local 
dissipation. It should be noted that this is not 
always the case [\cite{ortolani93,ji95}], but the assumption serves
to simply illustrate the key concepts for the present purposes,
and facilitates straightforward testing with spectral
code numerical simulations.
The time-evolution equations of magnetic helicity (e.g. [\cite{berger}]),
magnetic energy  and fluid kinetic energy  
for incompressible, resistive, viscous  MHD can then be written
\beq
\partial_t H\equiv 
\partial_t\lb\bfA\cdot \bfB\rb=-2\lambda\lb{\bf J}\cdot\bfB\rb,
\ee
\beq
\partial_t M\equiv\partial_t \lb\bfB^2\rb=
-2\lambda\lb(\nabla\bfB)^2\rb -2\lb {\bf v}\cdot({\bf J}\ts {\bf B})\rb,
\label{enold}
\ee
and 
\beq
\partial_t K\equiv\partial_t \lb\bfv^2\rb=
-2\nu\lb(\nabla\bfv)^2\rb +2\lb {\bf v}\cdot({\bf J}\ts {\bf B})\rb,
\label{enold2}
\ee
where $H$ is the magnetic helicity, $M$ is twice the magnetic
energy, $K$ is twice the kinetic energy, 
$\lambda$ is the
magnetic diffusivity, $\nu$ is the
viscosity, 
$\bfv$ is the velocity,  
and $\bf B$ is the magnetic field in Alfv\'en
units.  The vector potential $\bfA$ 
satisfies $\bfB\equiv \curl \bfA$, 
and the current density is normalized such that ${\bf J}=\curl\bfB$.
The brackets indicate a volume average.

In general,  the velocity should be 
included dynamically, as is done in [\cite{blackmanfield03}].  
Here I consider a case in which the velocity term can be approximately
ignored in the energy dynamics. Deriving a condition for
this requires an analysis of the spectra, as discussed in the appendix.
Note that even if such a condition were not satisfied,  the
magnetic energy loss rates and subsequent comparisons to those of magnetic
helicity would still be of interest: 
they would specifically represent the resistive dissipation of magnetic energy,
which would  not necessarily be approximately equal to the total magnetic 
energy change rate.


Now write
$M= \int_{k_L}^{k_\lambda} M_k dk$ 
with 
\beq
M_k\equiv \int |{\tilde \bfB}({\bf k})|^2 k^2d\Omega_k
=\int k^2|{\tilde \bfA}({\bf k})|^2 k^2d\Omega_k
 \propto k^{-q},
\label{E}
\ee
where the tilde indicates the Fourier transform,
${\tilde \bfB}({\bf k})=i{\bf k}\ts {\tilde\bfA}({\bf k})$,
$k$ is the wavenumber magnitude, and 
$\Omega_k$ is the solid angle in wavenumber space.
The second relation in (\ref{E}) 
follows in the Coulomb gauge $\nabla\cdot\bfA=0$,
and the assumed power-law form of $M_k$ 
takes $q$ to be independent of $k$
between $k_L$ and $k_\lambda >> k_L$, 
the smallest and largest wave numbers of the system  
(e.g. the inverse of the system size, and
the resistive wavenumber respectively).
Similarly, we write the magnetic helicity spectrum 
\beq
H_k\equiv 
{1\over 2}\int [{\tilde\bfA}({\bf k})\cdot{\tilde{\bf B}}^*({\bf k})+
{\tilde\bfA}^*({\bf k})\cdot{\tilde{\bf B}}({\bf k})]
k^2d\Omega_k =M_k f(k)/k,
\label{5a}
\ee
where  
\beq
f(k)\propto k^{-s}
\label{f}
\ee
is the assumed form of the fractional magnetic helicity as a function of $k$,
with $s$ independent of $k$, and $*$ indicates complex conjugate.
In the Coulomb gauge,  $f$ is the fractional helicity  
where 
$|f|\le 1$ (realizability condition [\cite{fplm}]),  
and $|f|=1$ is the maximal helicity case.
(Ref. [\cite{b2002}] provides further  
discussion in the context of numerical simulations.)
We also write  the  current helicity spectrum 
$C_k=k M_k f(k)$. 
The above definitions imply 
\beq
\lb\bfA\cdot\bfB\rb =H= \int_{k_L}^{k_\lambda} H_k dk = \int_{k_L}^{k_\lambda} f(k) M_kk^{-1} dk, 
\ee
\beq
\lb\bfJ\cdot\bfB\rb =C =  \int_{k_L}^{k_\lambda} 
k^2 H_kdk= \int_{k_L}^{k_\lambda}  f(k) k M_k dk, 
\ee
\beq
\lb\bfB^2\rb = M= \int_{k_L}^{k_\lambda}  M_kdk, 
\ee 
and
\beq
\lb(\nabla\bfB)^2\rb \simeq  \int_{k_L}^{k_\lambda}  k^2M_kdk.
\ee
These  formulae  can be used to calculate the ratio of
the time scale of magnetic helicity dissipation
to that of magnetic energy dissipation.
In general, $f(k)$ could have both positive and
negative signs, but for simplicity, we consider an initial condition in which
$f(k)$ is positive for all $k$ within the integration bounds. 
Justification comes from 3-D numerical simulations of closed turbulent systems 
which show that unless kinetic helicity 
is injected, the magnetic helicity basically
maintains the same sign over the spectral range, though 
transfer between scales does occur [\cite{b2001,maronblackman}].

For the magnetic helicity dissipation time scale, we then have  
\beq
\tau_H = -H/\partial_tH = H/2\lambda C ={
\int_{k_L}^{k_\lambda} f(k) M_k k^{-1} dk \over 2 \lambda \int_{k_L}^{k_\lambda} 
f(k) k M_k dk },
\label{th}
\ee
while for the magnetic energy decay time scale (assuming  resistive decay) we have
\beq
\tau_{M}\simeq -M/\partial_t M 
\simeq -M/(\partial_t M)_{res} =
{\int_{k_L}^{k_\lambda}  M_kdk \over 
2\lambda \int_{k_L}^{k_\lambda} k^2 M_kdk 
}.
\label{et}
\ee
In Fig. 1, 
the ratio $R\equiv \tau_H/\tau_M$ is plotted as a function
of $q$ and $s$ using
(\ref{th}) and (\ref{et}), for four different values of $k_L/k_\lambda$.
The plots show a significant parameter regime where $R\le 1$.

\section{Discussion}

\subsection{Understanding specific regimes analytically}

To better understand Fig. 1, it is useful to 
further  study the ratio $R$ analytically.
Using (\ref{E}) and (\ref{f}) 
and assuming $q\ne 1,3$ and $q+s\ne  0, 2$,
we have
\beq
\tau_H={(2-q-s)[k^{-(q+s)}]^{k_\lambda}_{k_L}\over 2 (q+s)\lambda [k^{(2-q-s)}]^{k_\lambda}_{k_L}}
\label{th2}
\ee
and 
\beq
\tau_M= {(3-q)[k^{1-q}]^{k_\lambda}_{k_L}\over 2\lambda(1-q) 
[k^{3-q}]^{k_\lambda}_{k_L}},
\label{et2}
\ee
where the brackets denote taking the difference between
the quantity inside evaluated at $k_\lambda$ and $k_L$.
I now explore several different value ranges of 
$q$ and $s$ and compute $\tau_M$ and $\tau_H$ and 
combine the results to determine $R\equiv \tau_H/\tau_M$.

First consider  $\tau_H$:
For $0<q+s<2$, Eqn. (\ref{th2}) gives
\beq
\tau_H={(2-q-s) k_L^{-(q+s)} \over 2 (q+s)\lambda k_\lambda^{(2-q-s)}}
\label{th1}
\ee
while 
for  $q+s> 2$, 
\beq
\tau_H={(2-q-s) \over 2 (q+s)\lambda k_L^2}.
\label{th2}
\ee
Now consider $\tau_M$:
For $1 < q < 3$, Eqn. (\ref{et2}) gives
\beq
\tau_M= {(3-q)k_L^{1-q}\over 2\lambda(1-q) 
k_\lambda^{3-q}},
\label{te1}
\ee
while for $q > 3$, 
\beq
\tau_M= {(3-q)\over 2\lambda(1-q) 
k_L^{2}},
\label{te2}
\ee
and for 
$0 < q < 1$
\beq
\tau_M= {(3-q)\over 2\lambda(1-q) 
k_\lambda^{2}}.
\label{te3}
\ee

Combining (\ref{th1}) and (\ref{te1})
for $0 <q+s<2$ and $1 < q < 3$, we have 
\beq
R= Q(k_\lambda/k_L)^{1+s},
\label{thte1}
\ee
where $Q\equiv|{(2-q-s)(1-q)\over (q+s)(3-q)}|$.
Combining (\ref{th1}) and (\ref{te2}) 
for $0 <q+s<2$ and $q > 3$, we have
\beq
R= Q(k_\lambda/k_L)^{q+s-2}.
\label{thte2}
\ee
Combining (\ref{th1}) and (\ref{te3})
for $0 <q+s<2$ and $0< q < 1$, we have 
\beq
R= Q(k_\lambda/k_L)^{q+s}.
\label{thte3}
\ee
Combining (\ref{th2}) and (\ref{te1})
for $q+s> 2$ and $1< q < 3$, we have 
\beq
R= Q(k_\lambda/k_L)^{3-q}.
\label{thte4}
\ee
Combining (\ref{th2}) and (\ref{te2})
for $q+s>2$ and $ q > 3$, we have 
\beq
R= Q.
\label{thte5}
\ee
Finally, combining (\ref{th2}) and (\ref{te3}), 
we have for $q+s >2$ and $0< q < 1$ 
\beq
R= Q(k_\lambda/k_L)^{2}.
\label{thte6}
\ee
Three qualitatively different regimes are revealed by 
(\ref{thte1}-\ref{thte6}):
{\bf I}. $R < 1$: Eqns. (\ref{thte1}) 
and (\ref{thte2}) show cases in which the magnetic helicity
can actually decay more quickly than the magnetic energy.
Eqn. (\ref{thte1})
shows that for $s<-1$ and $0<s+q<2$ and $1<q<3$, $R< 1$ can arise. 
In addition,  Eqn. (\ref{thte2}) shows that for $0< q +s <2$ and 
$q>3$, $R<1 $ is guaranteed.
{\bf II}. $R \simeq 1$:
Eqn. (\ref{thte5}) shows that $R$ is independent 
of the ratio of wave numbers when $q+s>2$ and 
$q>3$. For this regime, $Q>1$ so that the magnetic 
helicity decays more slowly, but $Q$ can be close to 1, so $R\simeq 1$.
{\bf III}. $R>1$:
For the remaining cases of Eqn. (\ref{thte1}) for $s>-1$, and  for all 
regimes of $q,s$ when Eqns. (\ref{thte3}), (\ref{thte4})  and   (\ref{thte6})
are applicable, $\tau_H > \tau_M$ 
by a factor that depends on $k_\lambda/k_L$.   
These three regimes can be seen within the continuous surfaces of 
Fig. 1, where $R$ was calculated directly from 
(\ref{th}) and (\ref{et}).

{In regime III, magnetic energy decays faster than magnetic helicity, 
and thus will indeed relax toward the Taylor state [{3-5}].
As just derived, this regime arises when both 
the magnetic energy and fractional magnetic helicity decrease
with increasing $k$, and only when the magnetic energy 
spectrum is not too steep;  regime II rather than regime III 
applies for a wide range of $s>0$, when  $q>3$.  Note that 
in contrast to regimes II and III, 
regime I can only result from an initial spectrum in which the fractional
magnetic helicity increases with decreasing wave number. 
The realizability condition ($|f(k)\le 1|$) [\cite{fplm}] 
then further implies 
the system is not  maximally helical at all scales when case I applies.
This in turn implies that there would be scales at which the
kinetic energy can equal or dominate the magnetic energy.
An example that could qualify for regime I 
would be one in which the kinetic energy and magnetic energy
were equal at large scales, but the velocity falls off
more rapidly than the magnetic energy at smaller scales.
Whether or not a particular physical phenomenon produces 
such given spectrum is not addressed here.  
The main point is just to highlight that the slow resistive decay
is of magnetic helicity is not guaranteed 
in all circumstances.  

A subtlety occurs if  $f(k)$ 
were to  change rapidly in time (e.g. if its spectral
index $s$ changes with time).  The estimated $\tau_H$ could then be time
dependent in such a way that the initial reduction of 
$H$ by a factor of $1/e$ could
occur in a different parameter regime from the subsequent
time evolution of $H$. 
For example, $H$ could decay to a value of $1/e$ faster than
$M$ reduces by the same fraction so that $R<1$ initially,
but the subsequent evolution
to $H=0$ might be slower than $M$ if 
the change in $f(k)$ induces a regime change such that 
$R>1$ at late times.

In principle, the results above could be used 
to help diagnose 
the initial fractional magnetic helicity and magnetic energy
spectra for a system in which the decay time scales of these
quantities can be measured. 
The ratio $\tau_H/\tau_M$ can then be compared with unity and the regime  
I,II or III revealed.
The results can certainly be  tested with a 3-D numerical
experiment in which magnetic helicity and energy spectra
are imposed initially to be in either regime I,II, or III and
the time scale for the decay of magnetic energy or magnetic helicity
monitored, and a comparison of the resistive decay terms measured.

\subsection{Relation to previous work}

It is useful  
to discuss the implications herein in the context of Ref. 
[\cite{berger2}]. The results of the two approaches are 
compatible, but complementary.  Using the Schwartz inequality, 
Ref. [\cite{berger2}] derived an upper bound to the total helicity decay
rate as a function of the magnetic energy and magnetic energy decay 
rate, which was then used to determine 
how much magnetic helicity $\Delta H$ would decay in the time
scale that magnetic energy decays, or,  for a given amount 
of helicity, how long it would take to decay this amount of helicity.
No explicit dependence of these decay rates on the 
spectral indices was considered. In the specific example of a coronal
loop considered in Ref. [\cite{berger2}], the helicity decayed more slowly than magnetic
energy, which was consistent with that example reflecting
a more nearly maximal helicity case.
Indeed as discussed above, 
$\tau_H > \tau_M$
in regime III. But above it was revealed that 
$\tau_H=\tau_M$ in regime II and $\tau_H < \tau_M$ in regime I.
These latter two regimes were not explicitly considered in Ref. [3].

The existence of the different regimes 
suggests that the commonly stated concept that 
``magnetic helicity decays on a longer time scale
than magnetic energy'' is not universally applicable.
However, the modified statement ``magnetic helicity resistively 
decays on the same or 
on a longer time scale than magnetic energy'' 
is fully justified when based on a comparison of 
$H_{max}/\partial_t H$ 
(or $H_{max}/\partial_t H_{max}$) 
to $M/\partial_t M$, where $H_{max}$ is   
the maximum possible magnetic helicity (i.e. corresponding to $f(k)=1$
for all $k$) for a given total magnetic energy. 
In this case, region I does not apply, and only region 
II and III apply.
Note however, that a given system is not guaranteed to have maximal
helicity.

A comparison of $H_{max}/\partial_t H$ to $M/\partial_t M$
also follows from equation (15) of Ref. [\cite{berger2}],
the key equation relating the two decay rates using the Schwartz inequality.
Using the present notation, the content of that equation is
\beq
H_{max}/\partial_t H
\ge (M/\partial_t M)^{1/2},
\ee
which is fully  compatible with, 
but complementary to, the spectral considerations above. It does not
by itself directly reveal the role of the spectral shapes.

\section{Conclusions}

Assuming that boundary terms are small, 
the influence of the spectral shapes of magnetic helicity
and magnetic energy on 
the resistive decay time scales of total magnetic helicity
and magnetic energy was determined.
Regimes exist for which the magnetic helicity decays more slowly,
more quickly, or at the same rate as magnetic energy. 
The conventional regime for which 
$\tau_H/\tau_M > 1$ 
applies to a wide range of spectra,
but the existence of regimes for which 
$\tau_H/\tau_M=1$ and $\tau_H/\tau_M < 1$, 
while  implicitly understood in some treatments of
magnetic helicity evolution in laboratory devices 
[e.g. \cite{ortolani93}], are often not emphasized when the basics of
magnetic helicity vs. magnetic energy decay are introduced
[\cite{bellan00}].  When the magnetic helicity is nearly maximal, 
$\tau_H/\tau_M \ge 1$, with the equality applying only when
the magnetic energy spectrum is      steep.

Because I have assumed  the velocity term in (\ref{enold}) is small
(see also the appendix) care must be taken when evaluating the implications 
for general systems incurring dynamical Taylor relaxation
in which velocity fluctuations must be solved for dynamically.
Indeed Eq. (\ref{enold}) reveals that the velocity term
provides an additional way for magnetic energy to drain into kinetic
energy and subsequently dissipate through viscosity.
This can help justify the 
conventional intuition that magnetic helicity decays 
more slowly than magnetic energy under certain circumstances and it
also  plays an important dynamical role in  driving a system 
toward and away from the Taylor state [\cite{ortolani93,blackmanfield03}]. 
However, since the velocity term can be of either sign,
it may also induce a slower decay of magnetic energy 
depending on the situation.
In the presence of dynamically significant velocity fluctuations,
the comparison made of time scales in this paper can then
be simply thought of as 
$\tau_H/(\tau_M)_{res}$, where the denominator represents the magnetic 
energy loss time through resistive dissipation only.

\section*{Appendix:  On ignoring the velocity term in Eq. (\ref{enold})}

It is helpful to compare the relative magnitudes
of the terms on the right of (\ref{enold}) to motivate
ignoring the velocity term.
To do so, we write the kinetic energy as
\beq
E= \int_{k_L}^{k_\lambda} E_k dk,
\ee
where $E_k\propto k^{-q_V}$.
We first work on the second term on the right of (\ref{enold}).
Noting that $\lb\bfv\cdot (\bfJ\ts\bfB)\rb\le
\lb(\bfv^2)^{1/2}k^2\bfB^2\rb$.
and using the spectral scalings for
kinetic energy combined along with
$M_k\propto k^{-q}$ from (\ref{E}),  we obtain 
\beq
\lb\bfv\cdot (\bfJ\ts\bfB)\rb\lsim k_LB_L^2
v_L(k_L/k)^{2q+q_V-5\over 2},
\ee
so that for $q< {5-q_V\over 2}$ 
\beq
\lb\bfv\cdot (\bfJ\ts\bfB)\rb\sim k_LB_L^2v_L
(k_\nu/k_L)^{5-2q-q_V\over 2}
\label{gg}
\ee
while for $q> {5-q_V\over 2}$ 
\beq
\lb\bfv\cdot (\bfJ\ts\bfB)\rb\sim k_LB_L^2v_L.
\label{ggg}
\ee
In the above equations, the viscous cutoff wavenumber $k_\nu$
has entered because
I have assumed that the magnetic Prandtl number $Pr_M\equiv \nu/ \lambda >>1$:
In magnetically dominated systems such as coronae or laboratory
devices, 
$Pr_M\simeq
(T/1.4\ts 10^{5}{\rm K})(n/10^{16}{\rm cm^{-3}})^{-1}>> 1$ [\cite{balbus98}]. 
Thus the maximum wavenumber for which 
$\lb\bfv\cdot (\bfJ\ts\bfB)\rb$ remains finite is $k_\nu> k_\lambda$,
explaining the appearance of $k_\nu$.  

For the first term on the right of (\ref{enold}), 
we again use $M_k\propto k^{-q}$ as in (\ref{E}).
We also assume that the decay time at $k_{\lambda}$, the dissipation
scale, is approximately equal to the dynamical time at that scale, equivalently, 
\beq
\lambda k_\lambda^2 \sim k_\lambda B_{k_\lambda},
\ee
remembering that $B$ is in Alfv\'en units.
It follows from some straightforward algebra that
\beq
\lambda \lb (\nabla \bfB)^2 \rb \sim k_L B_{L}^3 R_M^{5-3q\over q+1},
=k_L B_{L}^3 (k_\lambda/k_L)^{5-3q\over 2},
\label{mag}
\ee
where the magnetic Reynolds number 
$R_M\equiv B_L/k_L\lambda$.  
In this approach, 
the velocity term in (\ref{enold}) can be dropped when
Eqn. (\ref{mag}) $>>$ Eqn. (\ref{gg}) or Eqn. (\ref{ggg}),
depending on the regime.
More specifically, this condition becomes
\beq
{B_L\over v_L}\left({k_L/k_\nu}\right)^{q-q_V\over 2}\left
({k_\lambda /k_\nu} \right)^{5-3q\over 2}>> 1
\label{c1}
\ee
when $q < {5-q_V\over 2}$,
and 
\beq
{B_L\over v_L}\left({k_\lambda /k_\nu} \right)^{5-3q\over 2}>> 1
\label{c2}
\ee
when $q > {5-q_V\over 2}$.
Note that the square root of the ratio of 
magnetic to kinetic energies which enters
the above relations is that at the outer scale, not the total,
so the total magnetic energy
can be larger than the total kinetic energy
even the two energies do not satisfy this 
ordering at any given scale.

To see what the conditions above would imply for a specific
example, consider $B_L\sim v_L$ and let $q_V=5/3$.
Then ${5-q_V\over 2}=5/3$, so for $q<5/3$, the condition (\ref{c1})
would always be satisfied in this case.  It is common
to find  steeper magnetic energy spectra when compared to kinetic energy
spectra in MHD turbulence [\cite{maronblackman,cho02,haugen03,maroncowley04}].





\bigskip
\ni
Acknowledgments: 
Thanks to the referees for their comments which led
a number of clarifying revisions.
Support from DOE grant DE-FG02-00ER54600 is acknowledged.


\enumerate

\bibitem{woltjer}
L. Woltjer,  Proc. Nat. Acad. Sci., {\bf 44} 489 (1958).

\bibitem{berger} 
 M.A. Berger  \& G.B. Field,  J. Fluid Mech. {\bf 147} 133 (1984).

\bibitem{berger2} 
 M.A. Berger,   Geophys. Astrophys. Fluid. Dyn. {\bf 30} 79 (1984).

\bibitem{taylor86}
J.B. Taylor, Reviews of Modern Physics, {\bf 58}, 741 (1986) 

\bibitem{ortolani93} 
S. Ortolani  \& D.D. Schnack, 
{\it Magnetohydrodynamics of Plasma Relaxation}
(World Scientific: Singapore, 1993)

\bibitem{bellan00}  P.M. Bellan, {\sl Spheromaks}, 
(Imperial College Press, London, 2000)

\bibitem{heyvaertspriest84} J. Heyvaerts \& 
E.R. Priest, A.\& Ap., 137, {\bf 63}, (1984)

\bibitem{koniglchoudhuri85}A. K\"onigl \& 
A.R. Choudhuri, ApJ, {\bf 289}, 173, (1985)

\bibitem{ji95} H Ji.,S. C. Prager, J. S. Sarff, 
PRL {\bf 74} 2945 (1995).

\bibitem{blackmanfield03} E.G. Blackman \& G.B. Field, 
submitted to Phys. Rev. Lett., (2003), astro-ph/0303354.

\bibitem{fplm} 
 U. Frisch, A. Pouquet, J. L\'eorat \& A. Mazure,  J. Fluid Mech. 
{\bf
68}, 769 (1975).

\bibitem{b2002}
A. Brandenburg, W. Dobler, K. Subramanian,
Astron.Nachr. {\bf 323},  99 (2002)

\bibitem{b2001}
A. Brandenburg, ApJ, {\bf 550}, 824 (2001)

\bibitem{maronblackman}
J. Maron \& E.G. Blackman, ApJL {\bf 566}, L41 (2002). 

\bibitem{balbus98}
S. A. Balbus \& J.F. Hawley, Rev. Mod. Phys., {\bf 70}, 1 (1998).

\bibitem{cho02} J. Cho, 
A. Lazarian,  \& E.T. Vishniac, 2002, ApJL, {\bf 566} L49 (2002).

\bibitem{haugen03} N. Haugen, A. Brandenburg, W. Dobler,
 ApJL, {\bf 597} L141 (2003).

\bibitem{maroncowley04} J. Maron \& S. Cowley, in press ApJ (2004).

\vfill
\eject

\vspace{-.1cm} \hbox to \hsize{ \hfill \epsfxsize8.5cm
\epsffile{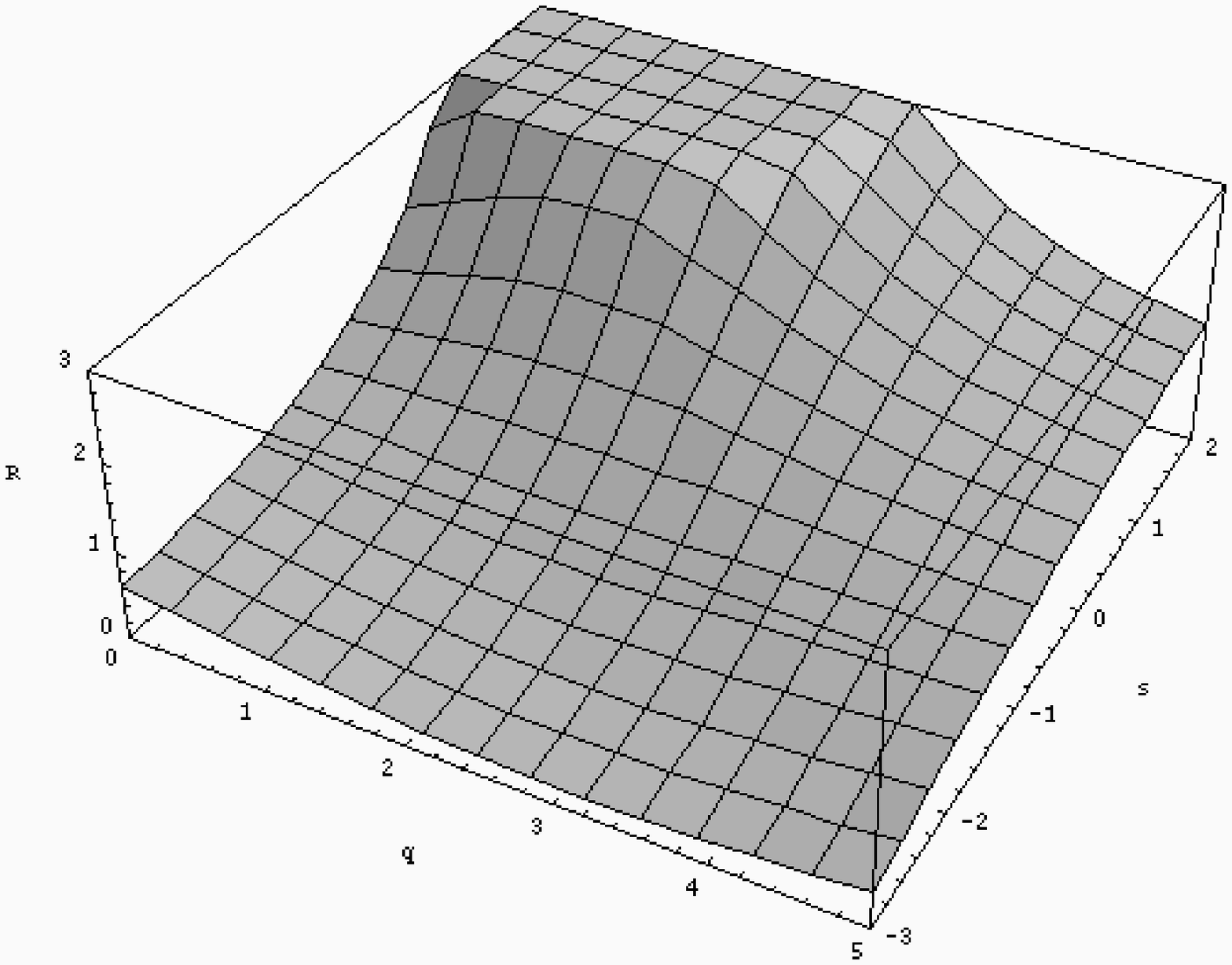} \epsfxsize8.5cm \epsffile{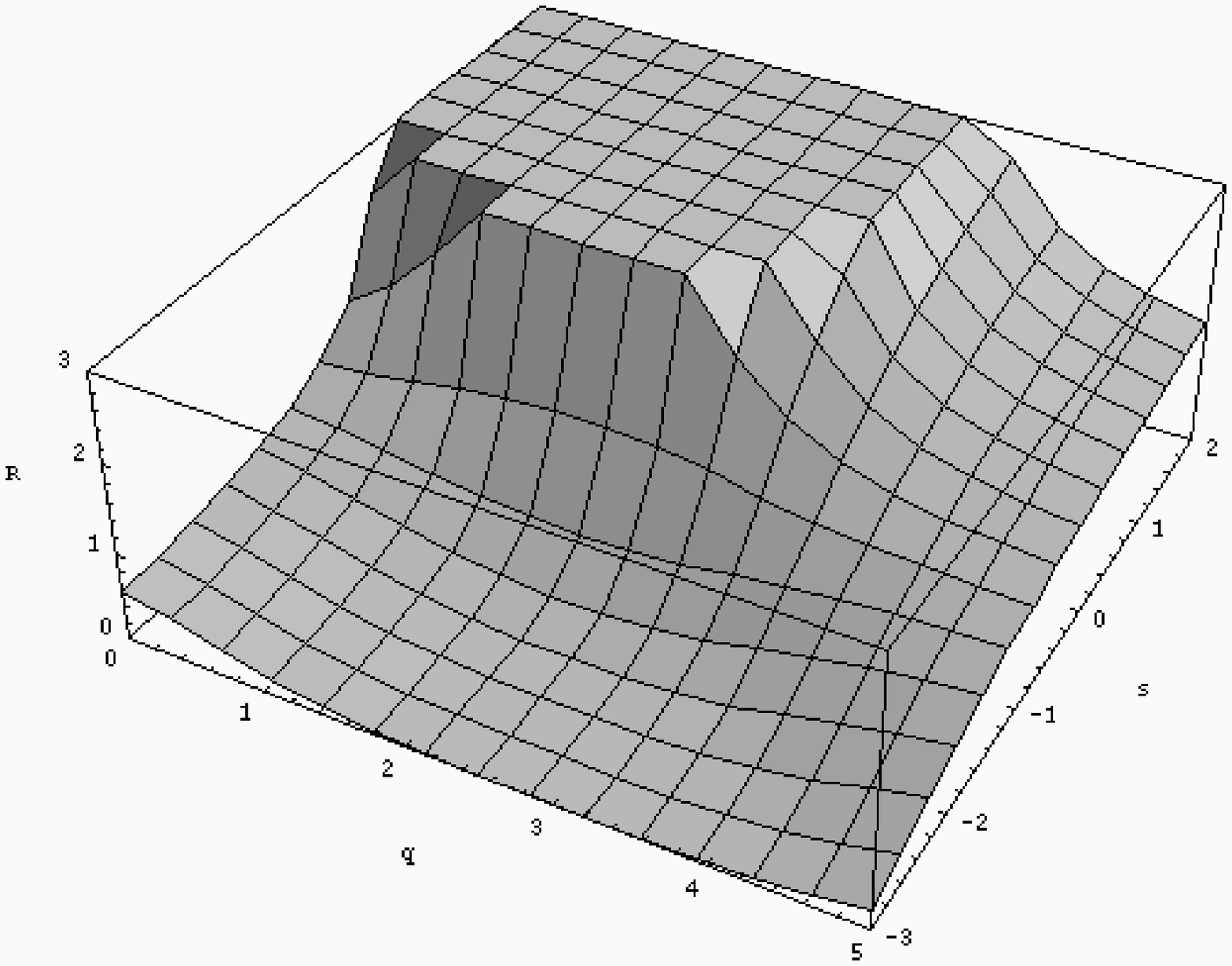} 
\hfill }
\vspace{-.1cm} \hbox to \hsize{ \hfill 
\epsfxsize8.5cm \epsffile{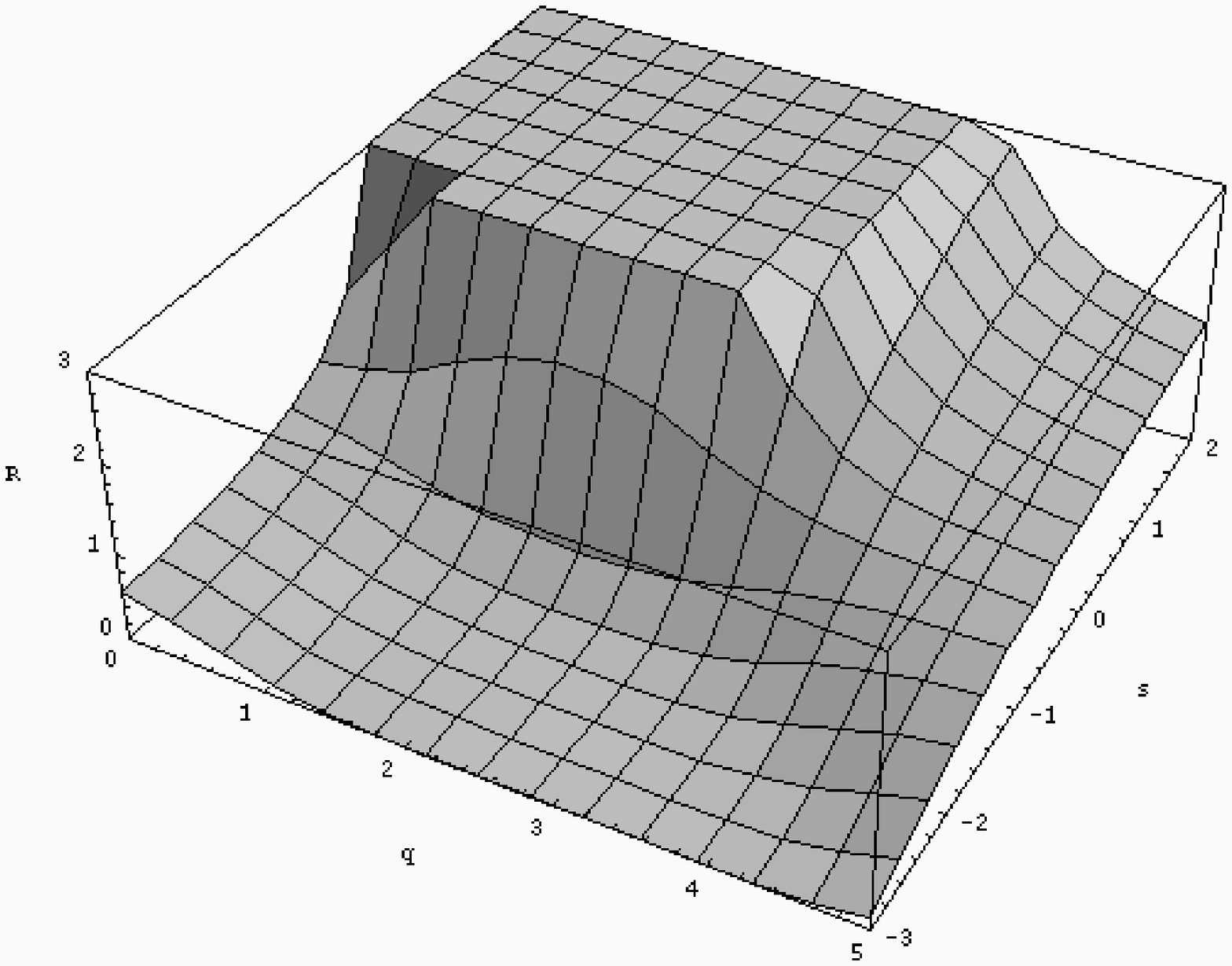} 
\epsfxsize8.5cm \epsffile {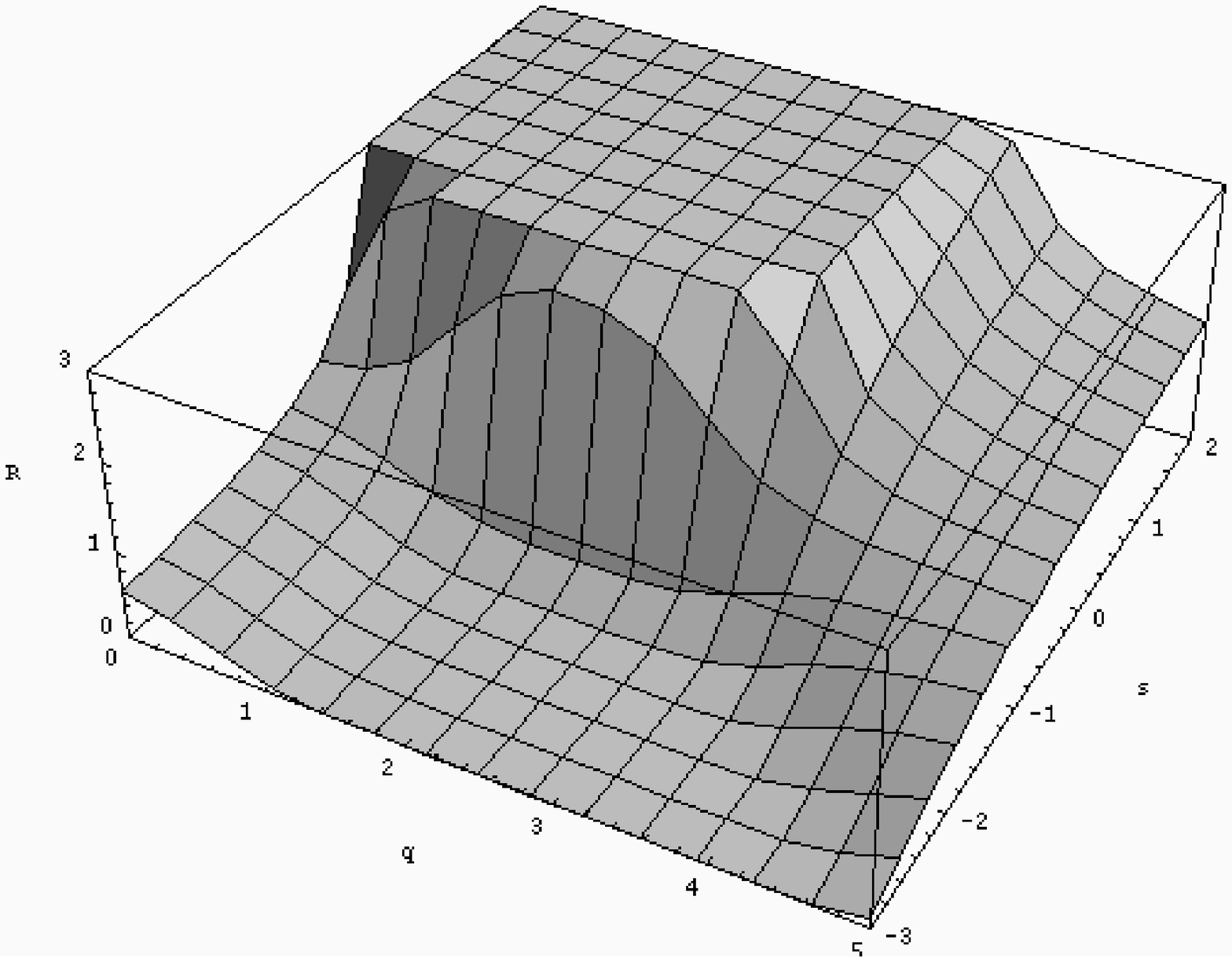} \hfill}

\noindent {
Figure 1: $R\equiv \tau_H/\tau_M$ vs. 
fractional helicty spectral index $s$, and energy spectral index $q$.
The plots were obtained by taking the ratio of (\ref{th}) to
(\ref{et}) for 4 different values of $k_\lambda/k_L$.
Clockwise from top left: $k_\lambda/k_L= 10, 10^2, 10^4, 10^3$.  
For $s>0$ and $3>q>0$, $R>>1$. This is the parameter regime III
in the text, and  corresponds to the commonly studied regime    
where magnetic helicity
decays more slowly than magnetic energy.  
However, note that for  small $q$ and $s<0$, $R<1$,   
and for $q > 3$, $R\lsim 1$. These correspond to regimes I and II
discussed below (\ref{thte6}).   The flat-top surfaces on each of the plots 
are just an artifact of 
truncating the vertical axis below the peak values of $R>>1$
attained by the actual solution surfaces.

\vfill
\eject
\end{document}